\documentclass[twocolumn, astrosymb, tighten]{aastex631}

\usepackage{amsmath}
\usepackage{mathtools}
\usepackage{url}

\usepackage{savesym}
\savesymbol{tablenum}
\usepackage{siunitx}
\restoresymbol{SIX}{tablenum}

\begin{document}

\title{Strong Nongravitational Accelerations and the Potential for Misidentification of Near-Earth Objects}
\shorttitle{Nongravitational Accelerations on NEOs}
\shortauthors{Taylor et al.}

\author[0000-0002-0140-4475]{Aster G. Taylor}
\altaffiliation{Fannie and John Hertz Foundation Fellow}
\affiliation{Dept. of Astronomy, University of Michigan, Ann Arbor, MI, 48109, USA}

\author[0000-0002-0726-6480]{Darryl Z. Seligman}
\altaffiliation{NSF Astronomy and Astrophysics Postdoctoral Fellow}
\affiliation{Dept. of Physics and Astronomy, Michigan State University, East Lansing, MI 48824, USA}
\affiliation{Dept. of Astronomy and Carl Sagan Institute, Cornell University, 122 Sciences Drive, Ithaca, NY, 14853, USA}

\author[0000-0002-1139-4880]{Matthew J. Holman}
\affiliation{Harvard-Smithsonian Center for Astrophysics, Minor Planet Center, 60 Garden St., Cambridge, MA, 02138, USA}

\author[0000-0002-5396-946X]{Peter Vere\v{s}}
\affiliation{Harvard-Smithsonian Center for Astrophysics, Minor Planet Center, 60 Garden St., Cambridge, MA, 02138, USA}

\author[0000-0003-0774-884X]{Davide Farnocchia}
\affiliation{Jet Propulsion Laboratory, California Institute of Technology, 4800 Oak Grove Dr., Pasadena, CA, 91109, USA}

\author[0000-0002-8507-1304]{Nikole Lewis}
\affiliation{Dept. of Astronomy and Carl Sagan Institute, Cornell University, 122 Sciences Drive, Ithaca, NY, 14853, USA}

\author[0000-0001-7895-8209]{Marco Micheli}
\affiliation{ESA NEO Coordination Centre, Largo Galileo Galilei 1, I-00044 Frascati (RM), Italy}

\author[0000-0001-6160-5888]{Jason T. Wright}
\affiliation{Dept. of Astronomy \& Astrophysics, The Pennsylvania State University, University Park, PA, 16802, USA}
\affiliation{Center for Exoplanets and Habitable Worlds, The Pennsylvania State University, University Park, PA, 16802, USA}
\affiliation{Penn State Extraterrestrial Intelligence Center, The Pennsylvania State University, University Park, PA, 16802, USA}

\correspondingauthor{Aster G. Taylor}
\email{agtaylor@umich.edu}

\begin{abstract}

Nongravitational accelerations in the absence of observed activity have recently been identified on NEOs, opening the question of the prevalence of anisotropic mass-loss in the near-Earth environment. {Motivated by the necessity of nongravitational accelerations to identify 2010 VL$_{65}$ and 2021 UA$_{12}$ as a single object, we investigate the problem of linking separate apparitions in the presence of nongravitational perturbations.} We find that nongravitational accelerations on the order of \qty{e-9}{\astronomicalunit/\day^2} can lead to a change in plane-of-sky positions of $\sim\qty{e3}{arcsec}$ between apparitions. Moreover, we inject synthetic tracklets of {hypothetical nongravitationally-accelerating} NEOs into the Minor Planet Center orbit identification algorithms. {We find that at large nongravitational accelerations ($|A_i|\geq\qty{e-8}{\astronomicalunit\per\day^2}$) these algorithms} fail to link a significant fraction of these tracklets. We {further} show that if orbits can be determined for both apparitions, {the tracklets} will be linked regardless of nongravitational accelerations, {although they may be linked to multiple objects}. {In order to aid in the identification and linkage of nongravitationally accelerating objects, we propose and test a new methodology to search for unlinked pairs. When applied to the current census of NEOs,} we recover {the} previously identified case but identify no new linkages. We conclude that {current linking algorithms are generally robust to} nongravitational accelerations, {but objects with large nongravitational accelerations may potentially be missed. While current algorithms are well-positioned for the anticipated increase in the census population from future survey missions, it may be possible to find objects with large nongravitational accelerations hidden in isolated tracklet pairs.} 

\end{abstract}


\section{Introduction} \label{sec:intro}

It has recently become evident that there exists a continuum of activity in small bodies in the solar system between the traditional definitions of comets and asteroids. Active asteroids \citep{Jewitt2012,Hsieh2017,Jewitt22_asteroid} and main belt comets  \citep{Elst1996,Boehnhardt1996,Toth2000,Hsieh2004,Hsieh2006} are objects on classically asteroid-like orbits that display visible activity in the form of faint comae. There also exist objects on cometary trajectories that appear inactive, such as ``Manx comets'' \citep{Meech2016} and  Damocloids \citep{Asher1994,Jewitt2005}. These intermediate objects could provide critical insights into our understanding of processes such as volatile and organic delivery to terrestrial planets \citep{Chyba1990,Owen1995,Albarede2009}, cometary fading \citep{Wang2014,Brasser2015}, and the depletion of volatiles or mantling on small bodies \citep{Podolak1985,Prialnik1988}. 

1I/`Oumuamua was the first interstellar object discovered traversing the inner solar system based on its hyperbolic trajectory \citep{Williams17}. {It} exhibited significant nongravitational acceleration in the radial direction of $A_1\sim\qty{2.5e-7}{\astronomicalunit/\day^2}$ \citep{Micheli2018}. However, deep optical imaging of the object displayed no dust coma \citep{Meech2017,Jewitt2017} or detection of carbon-based species outgassing \citep{Ye2017,trilling2018}. It is still unclear what mechanism is responsible for `Oumuamua's nongravitational acceleration, although \citet{Micheli2018} concluded that for realistic densities and geometries, outgassing with little or no associated dust production was the most likely cause. While its provenance remains unknown, \citet{Bergner2023} proposed that radiolytic production of H$_2$ from H$_2$O ice may be responsible for the acceleration. For recent reviews, see \citet{Fitzsimmons2023,Jewitt2023,Seligman2023_review}.

\citet{Chesley2016}, \citet{Farnocchia2023}, and \citet{Seligman2023} identified 7 near-Earth objects (NEOs) that also exhibited significant nongravitational accelerations and no associated dust coma (the ``dark comets''). Moreover, these accelerations were inconsistent with the typical causes of nongravitational acceleration on asteroids, such as the Yarkovsky effect \citep{Vokrouhlicky2015_ast4} or radiation pressure \citep{Vokrouhlicky2000}. Therefore, it was hypothesized that these objects were also outgassing with little dust production (at least when observed), similar to `Oumuamua and potentially due to seasonal effects \citep{Taylor2024}. 

A case of particular relevance in this context is that of 2010 VL$_{65}$.\footnote{\url{https://ssd.jpl.nasa.gov/tools/sbdb_lookup.html\#/?sstr=2010vl65}} \citet{Seligman2023} reported that the orbit of 2010 VL$_{65}$ could be readily linked with that of 2021 UA$_{12}$,\footnote{\url{https://ssd.jpl.nasa.gov/tools/sbdb_lookup.html\#/?sstr=2021ua12}} but only if an out-of-plane nongravitational acceleration of $A_3\sim\qty{9e-10}{\astronomicalunit/\day^2}$ was included in the fit. 

These examples show that compositional assumptions and/or the absence of observed activity cannot rule out nongravitational accelerations on objects in the near-Earth environment. Instead, astrometric orbit fitting is necessary to identify nongravitational accelerations for these objects. This methodology requires data over long arcs and several apparitions. If only gravity-only orbits are considered, nongravitational accelerations may disrupt these data arcs by causing objects to be incorrectly identified across apparitions, thereby obscuring nongravitational accelerations in the near-Earth environment. 

Motivated by the case of 2010 VL$_{65}$, we demonstrate that a sufficiently large nongravitational acceleration causes the on-sky position of an object to shift between apparitions in comparison to a gravity-only orbit. We quantify this change in on-sky position from nongravitational perturbations for dark comet test cases in Sec. \ref{sec:KY}, a synthetic population of asteroids in Sec. \ref{sec:population}, and for the currently-known NEOs in Sec. \ref{sec:neo}. This effect could plausibly lead to linkage failures, which would cause a single object to be falsely identified as distinct objects across apparitions. Therefore, in Sec. \ref{sec:MPCcheck}, we assess the accuracy of the current Minor Planet Center (MPC) linking algorithms. Informed by these results, in Sec. \ref{sec:nongrav_search} we search for unlinked pairs in the current census of NEOs.  We discuss our results and their broader ramifications in Sec. \ref{sec:discussion}. 

\section{Dark Comet Test Cases}\label{sec:KY}

In this section, we present ephemerides of two dark comets (1998 KY$_{26}$ and 2003 RM) under the influence of nongravitational perturbations. These objects are representative members of two distinct populations --- 2003 RM is large ($R_N\sim\qty{230}{m}$) and is on an orbit with $a\sim\qty{3}{au}$ and $e\sim0.6$, while 1998 KY$_{26}$ has $R_N\sim\qty{15}{m}$, $a\sim\qty{1.2}{au}$, and $e\sim0.2$. 

We use the \texttt{ASSIST} code \citep{Holman2023}, which is an extension of the \texttt{REBOUND} N-body code \citep{rebound} and the \texttt{REBOUNDx} library \citep{reboundx}. \texttt{ASSIST} integrates test particles with the IAS15 integrator \citep{Rein2015}, using precomputed positions of the Sun, Moon, planets, and 16 massive asteroids with the JPL DE441 ephemeris \citep{Park2021, Farnocchia2021}. As a result, \texttt{ASSIST} is significantly more efficient than a classic N-body integrator at the precision needed for ephemerides, at least in a solar system context. 

The \texttt{ASSIST} package can incorporate nongravitational accelerations using the \cite{MarsdenV} formulation, which we implement in this paper. Specifically, the  nongravitational acceleration takes the form
\begin{equation}\label{eq:nongrav}
\boldsymbol{a}_{\rm NG} = \left( A_1 \hat{\mathbf r} + A_2 \hat{\mathbf t} + A_3 \hat{\mathbf n}\right) \, g(r)\,.
\end{equation}
In Eq. \eqref{eq:nongrav}, the $g(r)$ function captures the dependence of the H$_2$O activity on the heliocentric distance $r$. We set $g(r)=(r_0/r)^2$, where $r_0=\qty{1}{au}$. The unit vectors $\hat{\mathbf r}$, $\hat{\mathbf t}$, and $\hat{\mathbf n}$ correspond to the radial, transverse, and out-of-plane directions with respect to the object's Keplerian orbit. The free parameters $A_1$, $A_2$, and $A_3$ provide the magnitude of the corresponding nongravitational acceleration component.

We integrate the trajectories of 2003 RM and 1998 KY$_{26}$ with a range of nongravitational accelerations in order to quantify the change in on-sky position. These simulations are initialized on 2000 January 1 and integrated for \qty{50}{yr}. For each object, we add a nongravitational acceleration in the form of Eq. \eqref{eq:nongrav}. We consider each component direction separately and a range of nongravitational acceleration magnitudes. For each nongravitational acceleration and direction, we calculate the change in the on-sky position $\Delta\Theta$ {relative} to the {trajectory} with the same initial conditions {and} gravity-only motion. The position change $\Delta\Theta$ is defined to be the offset angle between the on-sky positions of each trajectory. {If the geocentric direction vector is $\boldsymbol{n}_{\rm GO}$ on a gravity-only trajectory and $\boldsymbol{n}_{\rm NG}$ on a nongravitationally-accelerating trajectory, then} the difference in on-sky position is
\begin{equation}\label{eq:dtheta}
    \Delta\Theta=\cos^{-1}\left(\boldsymbol{n}_0\cdot\boldsymbol{n}_{\rm NG}\right)\,.
\end{equation}
Larger values of the $\Delta\Theta$ parameter correspond to larger discrepancies between the predicted (gravity-only) and perturbed on-sky position of the object. This could lead to misidentification of objects across apparitions --- a new observation may be categorized as a previously unidentified object rather than as an apparition of an already-known object. 

\begin{figure}
    \centering
    \includegraphics{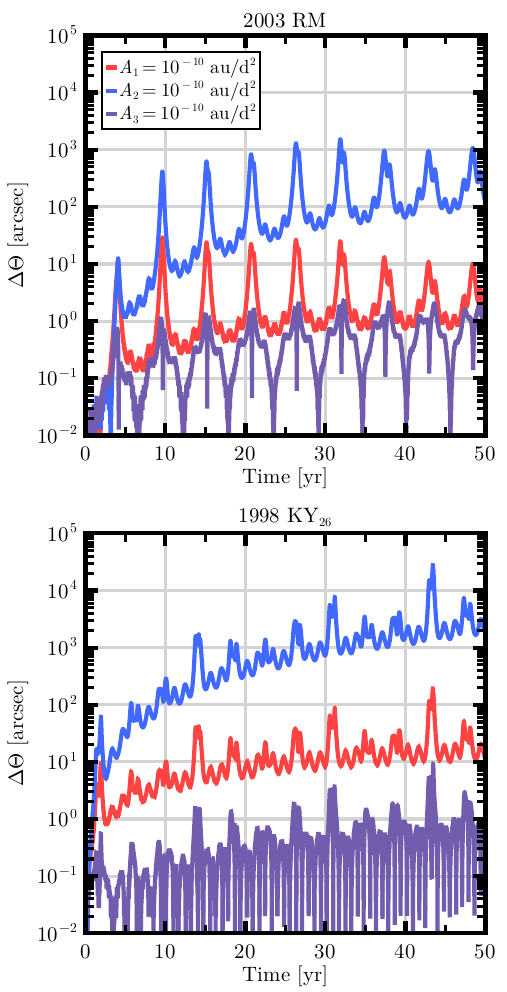}
    \caption{The change in the on-sky position (as seen from Earth) of 2003 RM (top) and 1998 KY$_{26}$ (bottom) as a function of time {for the three} nongravitational acceleration components. The magnitude of the {nongravitational} acceleration is {set to} $|A_i|=\qty{e-10}{\astronomicalunit/\day^2}$ and we show the radial $A_1$ (red), transverse $A_2$ (blue), and out-of-plane $A_3$ components (purple).  Since the ephemerides are linear in the nongravitational acceleration magnitude, stronger accelerations have the same functional form. The nongravitational accelerations are of the form given in Eq. \eqref{eq:nongrav}, and the position change $\Delta \Theta$ between the perturbed and unperturbed orbits is calculated using Eq. \eqref{eq:dtheta}.  }
    \label{fig:residuals_ky}
\end{figure}

\begin{figure*}
    \centering
    \includegraphics{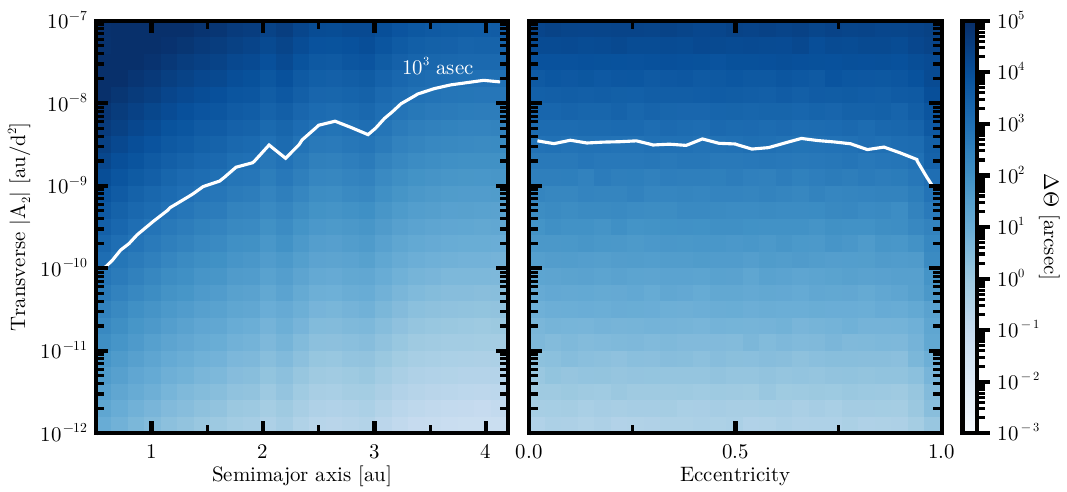}
    \caption{The median change in the on-sky position $\Delta\Theta$ after \qty{10}{yr} of integration as a function of the transverse nongravitational acceleration magnitude A$_2$, {semimajor axis, and eccentricity}. Each grid cell shows the {median} position change of all objects in a given bin. The white contour corresponds to  $\Delta\Theta=10^3$ arcsec.}
    \label{fig:dtheta_population}
\end{figure*}

In Fig. \ref{fig:residuals_ky}, we show the on-sky position change for all three nongravitational acceleration directions for 2003 RM and 1998 KY$_{26}$. We only show the orbits for a nongravitational acceleration magnitude $|A_i|=\qty{e-10}{\astronomicalunit/\day^2}$, but the small nongravitational acceleration magnitudes means that $\Delta\Theta$ {is a linear function of} acceleration magnitude \citep{Farnocchia2015}. This is only the case if the nongravitational acceleration is small compared to the gravitational acceleration due to the Sun, which takes the form of $A_1=\qty{-3e-4}{\astronomicalunit/\day^2}$ at 1 au. {In addition, if an object has a close encounter with a planet, then the ephemerides will no longer be linear. Due to the dynamically chaotic environment, position differences from nongravitational accelerations will lead to significant differences in the trajectories.}

It is evident that the nongravitational accelerations can drive significant changes in the on-sky position of both objects --- $\Delta\Theta\gtrsim\qty{10}{arcsec}$ --- over timescales comparable to the orbital period and apparition timescale. Moreover, the transverse nongravitational acceleration $A_2$ is far more efficient at modulating the sky position than the other component directions. This is consistent with expectations --- $A_2$ causes residuals to grow quadratically in time, $A_1$ causes residuals to grow linearly in time, and $A_3$ is not cumulative. 

{Note that the on-sky position changes shown in Fig. \ref{fig:residuals_ky} vary periodically in time in addition to the expected growth. The position change peaks during close approaches to Earth, since $\Delta\Theta\propto d^{-2}$, where $d$ is the distance between the observer and the object. The object is brightest and most likely to be observed during these close approaches. In addition, smaller objects are dimmer and are more likely to be observed only during close approaches, when the on-sky position changes are more significant. }

\section{Synthetic Population}\label{sec:population}

In this section, we generalize the results from Sec. \ref{sec:KY} to arbitrary orbits and accelerations. We {repeat our} ephemerides analysis on a synthetic population {that} spans the orbits of NEOs. 

This synthetic population consists of \num{e5} objects with orbital elements drawn from uniform distributions. Specifically, objects in the population are drawn with semimajor axis $a\in[0.5, 4.2)\,\unit{\astronomicalunit}$, eccentricity $e\in[0,1)$, inclination $i\in[0,15)\unit{\degree}$, argument of perihelion $\omega\in[0,360)\unit{\degree}$, longitude of ascending node $\Omega\in[0,360)\unit{\degree}$, and true anomaly at epoch $f_0\in[0,360)\unit{\degree}$. We then add nongravitational accelerations to each object and calculate {the change in on-sky position} relative to gravity-only orbits. We consider each of the nongravitational acceleration directions {and with} magnitudes logarithmic-uniformly sampled from $|A_i|\in[\num{e-12}, \num{e-7}]\,\unit{\astronomicalunit/\day^2}$. {After integrating for 1 and 10 \unit{yr},} we calculate the on-sky position changes. 

In Fig. \ref{fig:dtheta_population}, we show the median position changes after \qty{10}{yr} of integration time {under a transverse} nongravitational acceleration {versus semimajor axis and eccentricity}. This component is shown because it produces the largest on-sky position changes (see Sec. \ref{sec:KY}). Analogous calculations for (i) the radial and out-of-plane directions and (ii) shorter integration times are shown in Figs. \ref{fig:dtheta_A1}--\ref{fig:dtheta_A3_1year}. While $A_2$ generates the largest position change, it is evident from Figs. \ref{fig:dtheta_A1}--\ref{fig:dtheta_A3_1year}  that radial and out-of-plane accelerations can also drive significant position changes in some cases, even over shorter \qty{1}{yr} timescales. 

\begin{figure}
    \centering
    \includegraphics{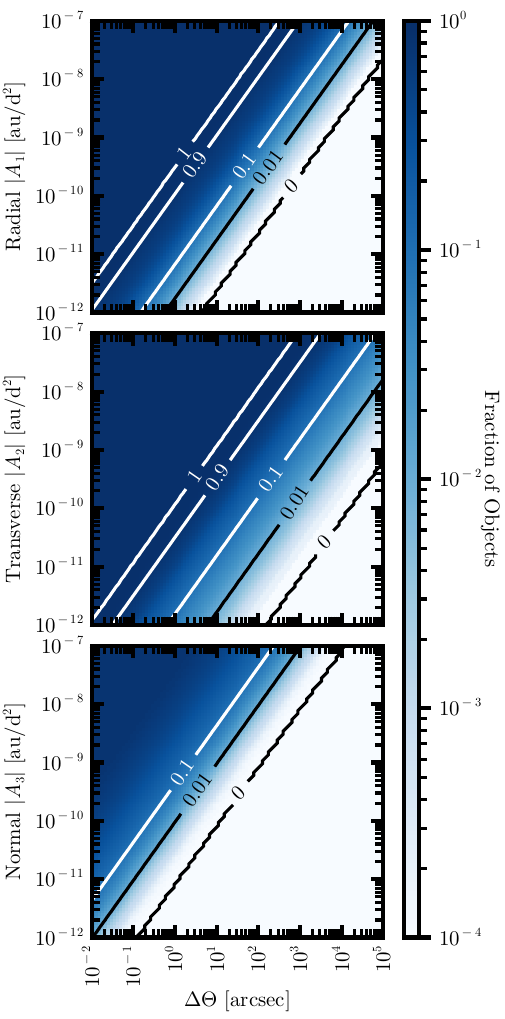}
    \caption{The fraction of NEOs that would exhibit a given change in sky position (or larger) over an approximate apparition period as a function of nongravitational acceleration magnitude and direction. The ephemerides are calculated for $|A_i|=\qty{e-10}{\astronomicalunit/\day^2}$ and linearly scaled for other nongravitational acceleration magnitudes.}
    \label{fig:CDF_allNEOS}
\end{figure}

\section{Application to NEO Population}\label{sec:neo}

In this section, we perform the same calculation as in Secs. \ref{sec:KY} and \ref{sec:population} for all currently-known NEOs.  For each NEO in the JPL Small-Body Database\footnote{https://ssd.jpl.nasa.gov/tools/sbdb\_lookup.html}, we calculate the on-sky position change for a general nongravitational acceleration. Each NEO is given a nongravitational acceleration with a magnitude of $|A_i|=\qty{e-10}{\astronomicalunit/\day^2}$ in each component direction. Each object is initialized at its orbital position when it was first discovered, and then integrated for an approximate apparition period. The apparition period is assumed to be {approximately} the larger of (i) the object's orbital period or (ii) the maximum gap between subsequent observations. The observational times are obtained from the MPC Small Bodies Database.\footnote{https://www.minorplanetcenter.net/data} 

After a single apparition period, we calculate the distance between the object's on-sky position with and without nongravitational accelerations. For a set nongravitational acceleration magnitude and direction, we calculate the fraction of all NEOS with corresponding on-sky position changes greater than a specified cutoff after an apparition. Although the integration is only performed for a single nongravitational acceleration magnitude, we scale the ephemerides by the nongravitational acceleration magnitude, since we assume that the ephemerides are linear (Sec. \ref{sec:KY}). We show these fractions in Fig. \ref{fig:CDF_allNEOS}.

Relatively weak nongravitational accelerations ($|A_2|\sim \qty{e-9}{\astronomicalunit/\day^2}$) can produce degree-scale on-sky position changes in $\gtrsim1\%$ of NEOs (Fig. \ref{fig:CDF_allNEOS}). This result should not be interpreted to mean that nongravitational accelerations exist on the calculated fraction of NEOs. Instead, these calculations are intended to suggest that NEOs with sufficiently strong nongravitational accelerations may have large changes in on-sky position.

\begin{table}
    \centering
    \caption{\textbf{Linkage Fraction.} The fraction of synthetic tracklets {(relative to gravity-only tracklets)} that are successfully linked by the MPC linkage algorithm for a given nongravitational acceleration magnitude and direction.}
    \begin{tabular}{c|lll}
        $|A_i|$ [\unit{\astronomicalunit/\day^2}] & Radial & Transverse & Normal \\\hline
         $10^{-12}$ & 0.9974 & 0.9947 & 1.0 \\
         $10^{-11}$ & 0.9974 & 0.9921 & 1.0026 \\
         $10^{-10}$ & 0.9921 & 0.9815 & 0.9974 \\
         $10^{-9}$ & 0.9868 & 0.9288 & 1.0106 \\
         $10^{-8}$ & 0.9525 & 0.3826 & 0.9736 \\
         $10^{-7}$ & 0.7467 & 0.0106 & 0.2876
    \end{tabular}
    \label{tab:linkfailfracs}
\end{table}

\section{MPC Linking Algorithm Fidelity}\label{sec:MPCcheck}

In this section, we present a preliminary proof-of-concept assessment of the fidelity of the MPC linking algorithm for NEOs with significant nongravitational accelerations. While the MPC also relies on user-reported linkages, the algorithms used by observers are varied and not publicly available. As a result, we only assess the fidelity of the MPC's internal {orbit fitting and} linkage pipeline. A detailed assessment of linking algorithm capabilities --- considering a comprehensive variety of nongravitational accelerations, orbits, apparition spacing, {and observation spacing} --- is warranted but is outside of the scope of this paper.

{The algorithm to} link observations across apparitions has two principle components --- attribution and identification. In the attribution process, novel astrometric observations are directly attributed to previously-known orbits. If the attribution process fails, it is still possible to link apparitions through the identification process. Once sufficient observations are made in the new apparition, a second orbit can be calculated and identified as {an extension of a} known orbit. We consider both of these processes independently, but note that both are used to link objects.

\subsection{Orbit Attribution}

The internal MPC attribution algorithm has two components --- \texttt{checkid} and \texttt{s9m}. The \texttt{checkid} algorithm takes the orbits of currently-known objects and {performs} a simple Keplerian advance to the date of the new data. It then calculates (i) the offset of the data from the predicted on-sky position and (ii) the change in the mean anomaly required to correct this offset. The \texttt{s9m} algorithm first reduces a given tracklet to the first and last observations contained within it. The position of the reduced tracklet is then compared to {the} predicted positions of the orbits of all previously identified objects. In contrast to \texttt{checkid}, the \texttt{s9m} algorithm propagates the fully perturbed orbits to the tracklet observations, although nongravitational accelerations are not included. {An object with a planetary close encounter between apparitions would be linked by \texttt{s9m}, but not \texttt{checkid}.} A linkage is made if (i) a previously identified  orbit is within the search radius {of 2 arcsec} and (ii) the on-sky velocity vectors of both objects are aligned. 

We assess the linkage fidelity of both of these algorithms for objects with a range of nongravitational acceleration magnitudes and directions. The sample contains \num{1000} currently-known NEOs that have multiple apparitions separated by less than \qty{15}{yr}.\footnote{The second criteria is imposed for the sake of computational efficiency.} For each object, we generate synthetic observations across an apparition, which is assumed to be the maximum separation between observations in current data. We first generate a short data arc, which consists of 2 observations separated by \qty{1}{\hour} on the 1st, 6th and 12th nights (6 total). The position of the object at the first observation time is obtained from the JPL Solar Systems Dynamics' Horizons System\footnote{https://ssd.jpl.nasa.gov/horizons/}. The positions at all other times are generated by numerical integration with \texttt{ASSIST}. We generate a tracklet at the subsequent apparition, which consists of 2 observations separated by \qty{1}{\hour} for 2 nights, for a total of 4 observations. {While such evenly-spaced observations are unrealistic, the observation spacing is kept constant for all objects. Therefore, the (generally dominant; \citealt{Milani2010}) errors from observation spacing are roughly constant for the different trajectories. } 

For each NEO, we generate tracklets with (i) a gravity-only orbit and (ii) {logarithmically-spaced} nongravitational accelerations in each component direction and magnitudes ranging from \qty{e-12}{\astronomicalunit/\day^2} to \qty{e-7}{\astronomicalunit/\day^2} (19 total tracklets for each object). Each synthetic observation accounts for the light travel time and reports the geocentric right ascension, declination, and apparent magnitude. Object apparent magnitudes are computed following the magnitude law for asteroids\footnote{Minor Planet Circular 10193} (not shown) with the absolute magnitude $H$ given by the JPL Small-Body Database and slope $G=0.15$, typical for asteroids.

We assess the fidelity with which the current MPC pipeline can link these synthetic tracklets to prior short-arc observations as a function of magnitude and direction of nongravitational acceleration. Due to the relatively sparse synthetic data, only $N_{\rm GO}=\num{379}$ of the \num{1000} gravity-only tracklets are successfully linked to their original objects by either \texttt{checkid} or \texttt{s9m}. {We calculate the number of tracklets $N_{\rm NG}$ that are successfully linked to prior short-arc data for each nongravitational acceleration magnitude and direction.} In Table \ref{tab:linkfailfracs}, we report {$N_{\rm NG}/N_{\rm GO}$,} the {fraction of nongravitationally-accelerating tracklets} that are successfully linked across apparitions, {relative to gravity-only trajectories.} 

The linkage fraction is greater than 1 for some out-of-plane nongravitational accelerations. This occurs because fitting an orbit to a tracklet is a slightly random process that depends on the precise on-sky position. At low acceleration magnitudes, out-of-plane nongravitational accelerations produce small changes in the on-sky position of a tracklet. These variations are not sufficient to cause a linkage failure but will induce small random variations in the orbit fitting to the tracklet. A few objects whose gravity-only tracklets failed orbit fits will be successfully fit and linked once nongravitational accelerations are included, leading to fractions that are slightly greater than 1. This does not indicate that nongravitational accelerations improve the linkage process, but that the linkage process is inherently stochastic. 

Successful linkage fractions significantly decrease in the presence of nongravitational accelerations with magnitudes larger than those identified for dark comets ($\gtrsim\qty{e-9}{\astronomicalunit/\day^2}$). Transverse nongravitational accelerations cause the sharpest decline in linkage fraction, and only $\sim90\%$ of tracklets with a transverse acceleration magnitude of $A_2\sim\qty{e-9}{\astronomicalunit/\day^2}$ are successfully linked. At a nongravitational acceleration magnitude of $A_2\sim\qty{e-7}{\astronomicalunit/\day^2}$, the linked fraction drops to $\sim1\%$. 

{While there is no known mechanism to produce such large transverse nongravitational accelerations (besides cometary outgassing), these results show that objects with these accelerations would be difficult to link and detect. However, radial nongravitational accelerations of this magnitude are known to exist and cause linkage failures in $\sim25\%$ of the considered objects. Therefore, it is possible that hypothetical NEOs with large nongravitational accelerations would not be correctly attributed across apparitions.}

\subsection{Orbit Identification}

It is still possible to link apparitions via orbit identification if attribution fails. If a sufficient number of observations are obtained during a new apparition, an independent orbit can be computed. These newly computed orbital elements are then compared to previously known orbits. If the two sets of orbital elements are within $3\sigma$ of each other, then the apparitions are linked. In this subsection, we test the fidelity of this process in the presence of nongravitational accelerations. {Specifically, we test if nongravitational accelerations can cause a previously-successful linkage to fail.}

We construct a sample of 214 NEOs that were successfully recovered across multiple apparitions from the MPC dataset. For each object, we identify the first apparition and record the times of each subsequent observation. We then calculate synthetic trajectories with a range of nongravitational accelerations and record the geocentric right ascension and declination at each observation time. Orbits are then individually determined for both the first and second apparition and the two sets of orbital elements are compared. The two apparitions are considered successfully linked if the two sets of orbital elements are separated by $<3\sigma$. 

The fraction of orbits that are successfully linked (generally $\sim80\%$) is independent of the magnitude of the nongravitational acceleration, {so nongravitational accelerations do not induce a linkage failure}. However, nongravitational accelerations cause issues in the orbit identification process. For sufficiently large nongravitational accelerations (generally $A\sim\qty{e-6}{\astronomicalunit/\day^2}$), the initial orbit determination will consistently fail, especially for long-arc data. This will prevent accurate linkage, since short-arc data will still be consistent with a gravity-only orbit. 

Moreover, the perturbations from nongravitational accelerations within each apparition tend to increase the uncertainty in the orbital elements calculated. In fact, the uncertainties for {nongravitationally-accelerating} orbits can be larger than the uncertainties for a gravity-only orbit by an order of magnitude or more. Since the orbit identification process uses a $3\sigma$ criteria, these large uncertainties increase the probability of a correct linkage. However, in this test we only compare {the apparitions of already-linked known objects, and ignore extraneous linkages to alternate objects.} In a realistic scenario, an orbit fit to a new apparition will be compared to all previously identified orbits. In this case, the larger uncertainties make it more likely that more than one {previously}-known object will be linked to the new apparition, confusing the identification. 

Even once a multiple linkage occurs, more observations will continue to refine the orbital parameter estimates. As the uncertainty shrinks, possible linkages will be pruned away until only one remains. However, the nongravitational accelerations will have modified the orbital parameters between apparitions. If the uncertainties shrink quickly, then {it is possible that a multiply-linked object will become unlinked and registered as new object}. This is likely what occurred for 2010 VL$_{65}$. {While false linkages to a large number of objects are possible, objects with the large nongravitational accelerations necessary to produce this effect will be rare and are unlikely to dramatically affect the NEO census. }

\section{Search for Unlinked NEOs}\label{sec:nongrav_search}

{In this section,} we search for unlinked pairs within the current census of NEOs. We consider the (i) orbital angular momentum, (ii) orbital elements, and (iii) novel {constants} of motion (derived in Sec. \ref{subsec:constant}) as metrics to link pairs of NEOs. {While} these metrics do not identify any unlinked pairs in the current NEO population, {they may be useful} for linking objects with nongravitational accelerations in future observations {or between isolated tracklets}. 

\subsection{Constants of Motion Under the Action of Nongravitational Accelerations}\label{subsec:constant}

{In this subsection, we derive quantities that are conserved under the action of nongravitational acceleration for a two-body system. These quantities provide a potentially useful metric for identifying unlinked pairs in the NEO population.}

{Consider an orbit characterized by the Keplerian orbital element set $\boldsymbol{x}=[a,e,i,\Omega,\omega,M_0]$. Under the action of nongravitational accelerations, these will have an orbit-averaged rate of change $\dot{\boldsymbol{x}}$. We wish to find a scalar function $F(\boldsymbol{x})$ that is constant  over the trajectories defined by $\dot{\boldsymbol{x}}$. Equivalently, the trajectories must be isosurfaces of $F$, or $\dot{\boldsymbol{x}}\cdot\nabla F=0$. }

{The rate of change of this phase-space position is given by }
\begin{equation}\label{eq:xdot}
    \dot{\boldsymbol{x}}=\frac{hr_0^2}{\mu a^2\eta}\begin{bmatrix}
                            2A_2a/\eta^2\\
                            A_2(1-\eta)/e\\
                            -A_3\cos\omega(1-\eta)/(e\eta)\\
                            -A_3\csc i\sin\omega(1-\eta)/(e\eta)\\
                            A_3\cot i\sin\omega (1-\eta)/(e\eta)\\
                            -2A_1\,.
                            \end{bmatrix}\,.
\end{equation}
{The quantity $\eta$ is defined to be $\eta=\sqrt{1-e^2}$. Eq. \eqref{eq:xdot} is derived by averaging the Gauss planetary equations \citep{Murray2000} over a single orbital period, assuming that the parameters do not change significantly over the period, or equivalently that the nongravitational acceleration magnitude is small. We additionally assume that the radial dependence of the nongravitational acceleration $g(r)\propto r^{-2}$, although analogous equations can be derived for a generic radial dependence. A derivation of these equations can be found in Appendix A.3 of \citet{Taylor2024}.}

{Our requirement that $\dot{\boldsymbol{x}}\cdot\nabla F=0$ becomes}
\begin{equation}\label{eq:longUdef}
\begin{split}
    0=&\frac{2a}{\eta^2}A_2\frac{\partial F}{\partial a}+\frac{1-\eta}{e}A_2\frac{\partial F}{\partial e}\\
    &-\frac{1-\eta}{e\eta}\cos\omega A_3\frac{\partial F}{\partial i}-\frac{1-\eta}{e\eta}\frac{\sin\omega}{\sin i}A_3\frac{\partial F}{\partial \Omega}\\
    &+\frac{1-\eta}{e\eta}\cot i\sin\omega A_3\frac{\partial F}{\partial \omega}-2A_1\frac{\partial F}{\partial M_0}\,.
\end{split}
\end{equation}
{We require that Eq. \ref{eq:longUdef} is satisfied for any orbital parameters and that $F$ does not depend on $A_1$, $A_2$, or $A_3$. In order to ensure that Eq. \eqref{eq:longUdef} holds for all values of $A_i$, we can divide $F$ into several functions $U$, $V$, and $W$ such that}
\begin{subequations}
\begin{align}
    0=&\frac{2a}{\eta^2}A_2\frac{\partial U}{\partial a}+\frac{1-\eta}{e}A_2\frac{\partial U}{\partial e}\,,\label{eq:Udef}\\
    \begin{split}
        0=&-\frac{1-\eta}{e\eta^2}\cos\omega A_3\frac{\partial V}{\partial i}-\frac{1-\eta}{e\eta}\frac{\sin\omega}{\sin i}A_3\frac{\partial V}{\partial \Omega}\\
        &+\frac{1-\eta}{e\eta}\cot i\sin\omega A_3\frac{\partial V}{\partial \omega}\,\text{, and}
    \end{split}\label{eq:Vdef}\\
    0=&-2A_1\frac{\partial W}{\partial M_0}\,.\label{eq:Wdef}
\end{align}
\end{subequations}
{We will solve each in turn.}

{Eq. \eqref{eq:Udef} can be solved by separation of variables to find that }
\begin{equation}\label{eq:U}
    U(a,e)=\ln\left[\frac{(\eta-1)^2}{a\eta^2}\right]\,.
\end{equation}
{Note that any arbitrary differentiable function $\mathbf{c}_1(U)$ is also a constant of motion.}

{Dividing Eq. \eqref{eq:Vdef} by $(1-\eta)A_3/(e\eta^2)$, we are left with the equation}
\begin{equation}
    0=\cos\omega\frac{\partial V}{\partial i}+\sin\omega\csc i\frac{\partial V}{\partial\Omega}-\cot i\sin\omega\frac{\partial V}{\partial \omega}\,.
\end{equation}
{This equation is not immediately separable. However, none of the terms depend on $\Omega$. As a result, we separate $V(i,\Omega,\omega)=G(i,\omega)K(\Omega)$. For a constant $\lambda$, we find that }
\begin{subequations}
\begin{align}
    K(\Omega)=&\exp(-\lambda \Omega)\,\\
    \lambda G(i,\omega)=&\sin i\cot\omega \frac{\partial G}{\partial i}-\cos i\frac{\partial G}{\partial \omega}\,.\label{eq:Gdef}
\end{align}
\end{subequations}
{Eq. \eqref{eq:Gdef} can further be solved to find that }
\begin{equation}
\begin{split}
    G(i,\omega)=&\mathbf{c}_2(2\sin i \sin\omega)\\
    \times&\exp(-\lambda\arctan(\cos i \tan\omega))\,,
\end{split}
\end{equation}
{where $\mathbf{c}_2$ is an arbitrary differentiable function. We select $\mathbf{c}_2(x)=\exp(x)$, set $\lambda=-1$, and note that $V(i, \Omega, \omega)=\mathbf{c}_3(K(\Omega)G(i,\omega))$. Taking $\mathbf{c}_3(x)=\ln(x)$, we find that }
\begin{equation}\label{eq:V}
    V(i,\Omega, \omega)=2\sin i \sin\omega+\arctan(\cos i \tan\omega)+\Omega\,.
\end{equation}

{The motion of a given object in orbital parameter space is constrained to follow isosurfaces} of Eqs. \eqref{eq:U} {and \eqref{eq:V}}. {Isosurfaces of the $U$ parameter  are 1-dimensional curves that will be followed precisely. However, isosurfaces of the $V$ parameter  are  2-dimensional manifolds, so the phase-space trajectory is not immediately given.} 

{The solution to Eq. \eqref{eq:Wdef} is trivial --- $W(M_0)=c$, where $c$ is a constant. This implies that no constant of motion can be constructed that depends on the mean anomaly at epoch $M_0$. }

{It is hypothetically possible to define another (linearly independent) scalar function $Q(i,\Omega,\omega)$ that is also a constant under nongravitational accelerations, which would provide the trajectory. However, if such a function exists, it must be the case that $\dot{\boldsymbol{x}}\cdot\nabla Q=0$. Since $\nabla V$ and $\nabla Q$ are both orthogonal to $\dot{\boldsymbol{x}}$, if $Q$ exists we must be able to define $Q$ and $V$ such that $\nabla V\times\nabla Q=\dot{\boldsymbol{x}}$. However, it can be shown that $\nabla\cdot(\nabla V\times\nabla Q)=0$ for all $V,Q$, while $\nabla\cdot\dot{\boldsymbol{x}}=-\cot i \cos\omega\neq0$. Therefore, while we have found a function $V$, no scalar function $Q$ exists.} \footnote{{It is possible to construct a scalar pseudodensity function $\rho$ such that $\nabla\cdot(\rho\,\dot{\boldsymbol{x}})=0$. Such a function enables the construction of $V$ and $Q$ such that $\nabla V\times\nabla Q=\rho\,\dot{\boldsymbol{x}}$. However, there is no closed-form expression for $\rho$, so it cannot be used to construct useful constants of motion.}}

\begin{figure}
    \centering
    \includegraphics{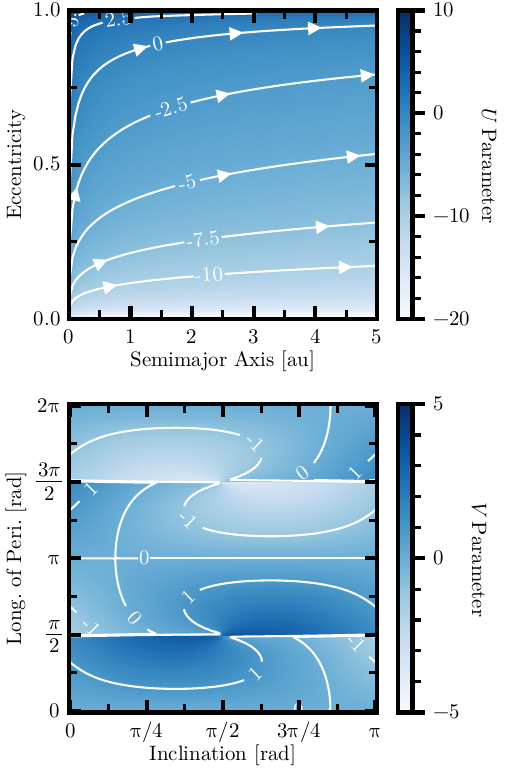}
    \caption{The $U$ parameter(Eq. \ref{eq:U}) over a range of ($a,e$) values {and the $V$ parameter (Eq. \ref{eq:V}) over a range of ($i,\omega$) values (we set $\Omega=0$). Contours are shown for both functions. For the $U$ parameter, the contours trace the phase-space trajectories of these objects, which is not the case for the $V$ parameter.} {In the top panel, } the arrows indicate the direction of motion for positive $A_2$. {The contours shown in the bottom panel are not equivalent to the phase-space trajectories. } }
    \label{fig:Uheatmap}
\end{figure}

The value of the $U$ {and $V$} parameters are shown in Fig. \ref{fig:Uheatmap}. {For the $U$ parameter, the contours of motion are also shown. }

In order to verify that {these quantities are} conserved, we numerically integrated test particles using \texttt{REBOUNDx} \citep{reboundx} for a variety of orbital parameters and calculated the value of the $U$ {and $V$ parameters} over time. For nongravitational accelerations small enough that the orbit-averaged equations of motion hold,\footnote{This is typically the case when $|A|\lesssim\qty{e-6}{\astronomicalunit/\day^2}$ for NEOs, although there is a weak dependence on the orbital period.} the $U$ {and $V$ parameters are} conserved to machine precision. In the presence of the planets, however, the orbital elements can be subjected to periodic oscillation due to secular perturbations. As a result, a single object may exhibit oscillations in {these parameters}, although these oscillations are only of magnitude \num{e-4}. 

\begin{figure}
    \centering
    \includegraphics{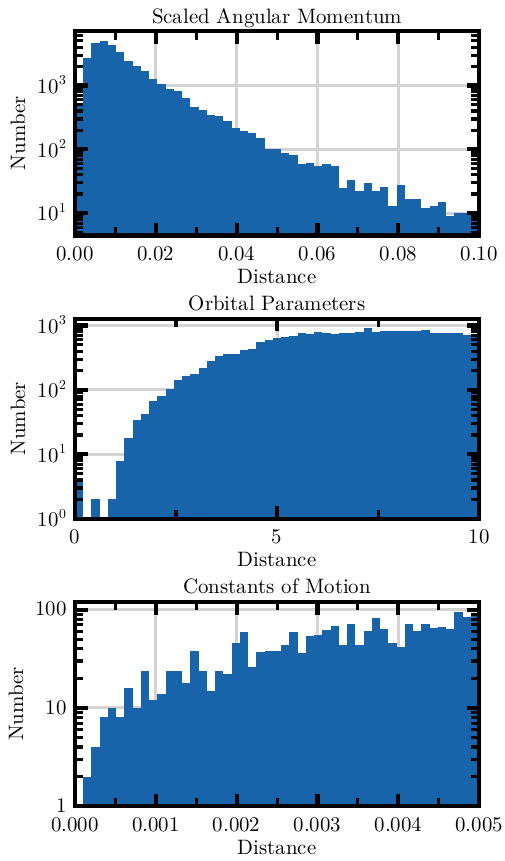}
    \caption{Histograms of the nearest-neighbor distance in orbital angular momentum (top),  orbital elements (middle), and the {constants} of motion given by Eq. \eqref{eq:U} {and Eq. \eqref{eq:V}} (bottom). }
    \label{fig:histograms}
\end{figure}

\subsection{Search Algorithms for Unlinked Pairs}\label{subsec:pairs_search}

In this subsection, we use the {constants} of motion derived in Sec. \ref{subsec:constant}, the orbital angular momentum, and the orbital elements to search for unlinked pairs in the current census of NEOs. Specifically, we calculate each quantity for each object in the MPC population of currently known NEOs described in Sec. \ref{sec:neo}. For each quantity, we construct a $k$-d tree for the data and extract the nearest neighbor in the $N$-dimensional parameter space. In Fig. \ref{fig:histograms}, we show histograms of the distance to the nearest neighbor for each NEO in the total population. 

We then obtained the closest 0.1\% of nearest-neighbor pairs for each parameter set. We used a least-squares fitting method to determine if these objects could be linked by a nongravitational acceleration (see \citealt{Farnocchia2015} for a more thorough description of the orbit-fitting algorithm). This methodology was able to recover the previously identified case for 2010 VL$_{65}$ and 2021 UA$_{12}$, which {was identified as a pair by all comparison metrics}. Although we recovered many potential pairs of unlinked objects, orbital fitting through object pairs did not recover a single linkage, even when accounting for possible nongravitational accelerations. We therefore conclude that each of these pairs are two distinct objects, with the exception of the previously-identified 2010 VL$_{65}$.  

We also tested the accuracy of this methodology for finding linked pairs. We added clones of each object {in the current population of NEOs} with varying levels of nongravitational acceleration and integrated for one orbital period. We assume that the nongravitational accelerations are sufficiently small that the change in orbital parameters is in the linear regime. We then calculated the fraction of the cloned objects that are {the} nearest neighbor {of} the original object {in the $k$-d tree}. For sufficiently large nongravitational accelerations, the cloned objects are distinct enough from the original that nearest-neighbor searches will not correctly match these objects. We find that the accuracy of these methods scales with the dimensionality of the parameter under consideration --- higher dimensions improves identification of the objects. {Regardless,} at a magnitude of \qty{e-6}{\astronomicalunit/\day^2}, only 88\% of cloned objects are linked by {comparing} orbital parameters. Therefore, more than 1 in 10 clones are nearer to another object than their original. For an acceleration with magnitude \qty{e-7}{\astronomicalunit/\day^2}, 99\% of cloned objects are {correctly identified}. Our search for pairs by orbital parameter implies that there are few, if any, unlinked pairs of NEOs with nongravitational accelerations below this magnitude. However, the assumption of linearity begins to break down for nongravitational accelerations of order \qty{e-6}{\astronomicalunit/\day^2} (when the acceleration is greater than 1\% of the solar acceleration). As a result, {hypothetical} objects {with stronger nongravitational accelerations would not be identified with the methodology presented in this section, although such objects are expected to be rare.} 

\section{Discussion}\label{sec:discussion}

In this paper, we demonstrated that nongravitational accelerations can significantly alter the on-sky positions of small bodies. Specifically, we showed that accelerations with magnitudes comparable to or greater than those measured on the the currently-known dark comets produce on-sky position changes of $\Delta\Theta\sim\qty{e3}{arcsec}$ over 1--10 \unit{yr} apparition timescales. {However}, we showed that the algorithms currently in use for {fitting orbits and} linking objects across apparitions {are generally robust in the presence of nongravitational accelerations of reasonable magnitude, although objects with large nongravitational accelerations ($|A_i|\gtrsim\qty{e-8}{\astronomicalunit\per\day^2}$) may be missed. While these acceleration magnitudes are expected to be rare, our results show that NEOs with such a nongravitational acceleration would be difficult to identify. Therefore, this population is not necessarily well-characterized, and may be more common than currently anticipated.}

Notably, $\sim73\%$ of NEOs in the MPC database have data arcs that are less than a year, and $\sim55\%$ have data arcs shorter than a month. In Sec. \ref{sec:population}, we demonstrated that for nongravitational accelerations stronger than the dark comets' by an order of magnitude, $\sim$10\% of NEOs are offset from their predicted on-sky location by more than a degree. Note that this difference is in comparison to a precisely known gravity-only orbit, while actual astrometric positions will include nongravitational accelerations. As a result, the calculated orbits will have errors that may compensate for the residual from nongravitational acceleration, weakening the linkage error.   

In Sec. \ref{sec:MPCcheck}, we investigated the fidelity of the linking algorithm currently in use at the MPC. This process has two principal steps --- attribution and identification. In the attribution step, new observations are compared to the expected locations of previously-known objects. If the objects are sufficiently close, then the new observations are attributed to the prior object. We determined that transverse nongravitational accelerations an order of magnitude larger than those of the dark comets could lead to an attribution failure rate of $\sim10\%$. However, there are several caveats to this assessment. Our sample is restricted to objects with long data arcs observed across at least one apparition, and only $\sim30\%$ of objects could be successfully linked even in the absence of nongravitational accelerations. {In addition, objects with transverse nongravitational accelerations with magnitudes $|A_2|\geq\qty{e-8}{\astronomicalunit\per\day^2}$ are hypothetical with no known production mechanism. However, radial nongravitational accelerations of magnitude $|A_1|\simeq\qty{e-7}{\astronomicalunit\per\day}$, which are possible on comets, can cause linkage failure for $\sim25\%$ of considered objects.}

Even if an object is not correctly attributed, a linkage can still be made by the orbit identification process. Further observations are collected until a second orbit can be calculated. If these orbital elements are within $3\sigma$ of a previously-known orbit, then a linkage is made. We found that the orbit identification process is robust to nongravitational accelerations, as it begins to fail only when gravity-only orbits cannot be calculated for the object's trajectory. In the presence of nongravitational accelerations, however, {successful} orbit identification is primarily driven not by the similarity in orbital parameters but the large magnitudes of the uncertainties, which are significantly increased by nongravitational perturbations. As a result, many objects may fall within the $3\sigma$ cutoff and be linked to the new apparition, {confusing} the identification. This complication will likely be exacerbated as the census of solar system small bodies expands. Further investigation is required to determine the significance and prevalence of these issues.

{In addition to the on-sky position changes induced by nongravitational accelerations, trajectories will have aleatory uncertainties as the result of short data arcs \citep{Milani2010}. Such aleatory uncertainties in the ephemerides can often be larger than the uncertainties induced by nongravitational accelerations, contributing to the problem of linking observations. In this work, we compare successful linkages between gravity-only and nongravitationally accelerating trajectories, keeping observation times and data arcs constant. Therefore, our results control for these aleatory effects and depend only on the nongravitational accelerations. While a full investigation of the combined effects of aleatory and epistemic errors in linking is worthwhile, it is beyond the scope of this work.}

We also searched for unlinked object pairs using three metrics that should be approximately conserved across apparitions (Sec. \ref{sec:nongrav_search}). Although we successfully recovered previously linked objects, we did not identify any new unlinked pairs. However, a more rigorous search is warranted, given that larger nongravitational accelerations may cause our pair-identification metrics to fail. 

These NEO-pair metrics may be used to identify smaller-magnitude nongravitational accelerations with future observations. Forthcoming observatories such as the the Vera C. Rubin Observatory Legacy Survey of Space and Time (LSST, \citealt{Schwamb2023}) and \textit{NEO Surveyor} \citep{Mainzer_Wright2015, Wright2021, Mainzer2022} will expand the census of these objects. With an increased population of NEOs, it {may be important} to account for nongravitational accelerations when linking objects across apparitions. Moreover, these surveys will also discover smaller objects, which exhibit stronger nongravitational accelerations per unit mass-loss rate than large objects. {In addition, smaller objects are dimmer and are more likely to be observed only during close approaches, when the on-sky position changes are more significant.} Therefore, accounting for such nongravitational accelerations may modify the census of NEOs produced by these surveys and identify more volatile-rich bodies in the near-Earth environment. {It may also be possible to identify large nongravitational accelerations by mining the Isolated Tracklet File for unlinked observations using the methods discussed in Sec. \ref{sec:nongrav_search}.}

A larger population of hydrated small bodies in the near-Earth environment could have ramifications for current theories of terrestrial volatile delivery. The Earth's water content is generally expected to have been delivered, at least in part, by hydrated small bodies from the outer solar system instead of accretion in situ \citep{Walsh2011, Raymond2014, OBrien2014}. The Earth's isotopic deuterium to hydrogen ratio (D/H; amongst other isotopic ratios) is significantly higher than the protosolar nebula, but significantly lower than the long-period or Jupiter-family comets that were once thought to be the source of Earth's oceans \citep{Owen1995, Meech2020}. While MBCs match the D/H ratio relatively well \citep{Alexander2012, Marty2012, Hallis2015}, the origins of terrestrial water remain uncertain. For a more detailed review of the origins of Earth's water, see \citet{Meech2020}. If the NEO population contains more hydrated objects than anticipated, as suggested by the dark comets, these objects may have contributed to this delivery process. {The likely origin of the dark comets in the inner main belt further suggests that volatiles may be common in the near-Earth environment \citep{Taylor2024b}.} Our results imply that {it is possible (but unlikely) that more volatile-rich NEOs may exist, but remain undetected due to failures in linkage algorithms.}

Moreover, if the future census multiply-counts objects and fails to account for nongravitational accelerations, then orbit predictions and Earth-impact probabilities may be flawed. Since NEOs are the primary source of Earth impactors \citep{Strom2005}, accounting for nongravitational accelerations may be necessary to fully understand the near-Earth environment and possible future impacts. Specifically, it is possible that undetected nongravitational accelerations could cause a seemingly-innocuous NEO to become an Earth impactor. Therefore, our results may have significant implications for both planetary defense and volatile delivery for terrestrial planets. In the context of the upcoming expansion of the census of known NEOs, correctly accounting for nongravitational accelerations {may be} critical {for accurate orbital predictions.}

Our results require significant future work to further characterize {the issues discussed here}. Since the MPC relies heavily on observers to {fit orbits and} link objects, a survey of the accuracy of current and future linkage methodologies (i.e., HelioLinC; \citealt{Holman2018}) is warranted. The tests of the MPC linkage algorithms presented here are limited in scope and only intended to {probe} a variety of possible issues. Further quantification of the prevalence and severity of possible linking problems is necessary to clarify this issue. {Although the MPC's current algorithms appear robust,} the future inclusion of nongravitational accelerations in NEO linkage algorithms may overcome these problems and provide a more complete understanding of nongravitational accelerations in the near-Earth environment. 

\section*{Acknowledgements}

{We thank the anonymous reviewer for their helpful comments.} We thank Paul Abell, Adina Feinstein, Linda Glaser, Megan Mansfield,  Luis Salazar Manzano, Robert Jedicke, Ken Chambers, Karen Meech, {W. Garrett Levine,} Jonathan Lunine, Samantha Trumbo, and Jonathan Williams for useful discussions. A.G.T. acknowledges support by the Fannie and John Hertz Foundation and the University of Michigan's Rackham Merit Fellowship Program. D.Z.S. is supported by an NSF Astronomy and Astrophysics Postdoctoral Fellowship under award AST-2303553. This research award is partially funded by a generous gift of Charles Simonyi to the NSF Division of Astronomical Sciences. The award is made in recognition of significant contributions to Rubin Observatory’s Legacy Survey of Space and Time. Part of this research was carried out at the Jet Propulsion Laboratory, California Institute of Technology, under a contract with the National Aeronautics and Space Administration (80NM0018D0004). The Center for Exoplanets and Habitable Worlds and the Penn State Extraterrestrial Intelligence Center are supported by Penn State and its Eberly College of Science. 

This paper made use of the Julia programming language \citep{Julia}, the plotting package Makie \citep{Makie}, the N-body code \texttt{REBOUND} \citep{rebound}, and the ephemerides-quality integrator \texttt{ASSIST} \citep{Holman2023}. This research has made use of data and/or services provided by the International Astronomical Union's Minor Planet Center. 

\bibliography{main}{}
\bibliographystyle{aasjournal}

\appendix
\restartappendixnumbering

\section{Supplemental Material}\label{appendix:population}

In this appendix, we show the on-sky position changes for the synthetic population described in Sec. \ref{sec:population} for a range of integration times and nongravitational acceleration directions. 

\begin{figure*}[!htb]
    \centering
    \includegraphics{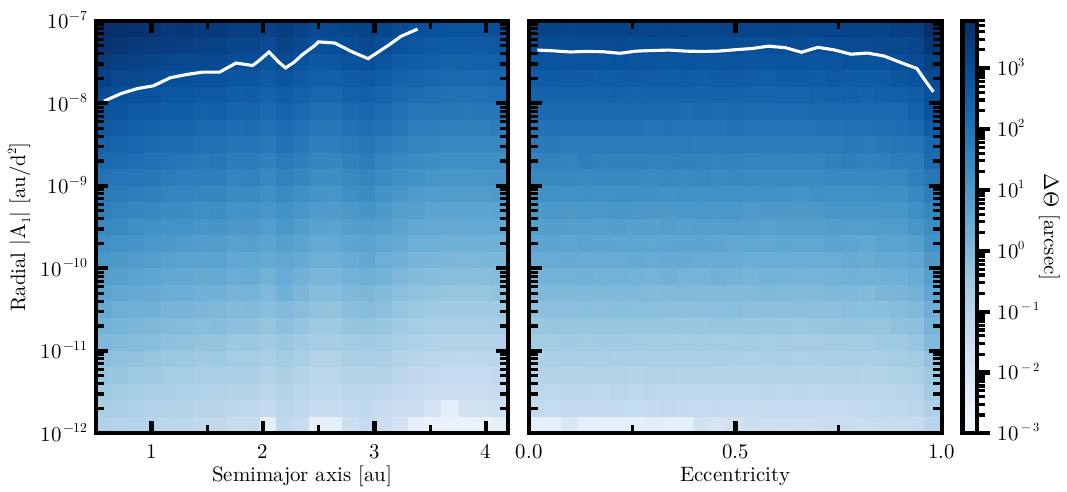}
    \vspace{-10pt}
    \caption{Similar to Fig. \ref{fig:dtheta_population}, but with radial A$_1$ acceleration instead of transverse A$_2$ acceleration. The contour is at \qty{e3}{arcsec}.}
    \label{fig:dtheta_A1}
\end{figure*}

\begin{figure*}[!htb]
    \centering
    \includegraphics{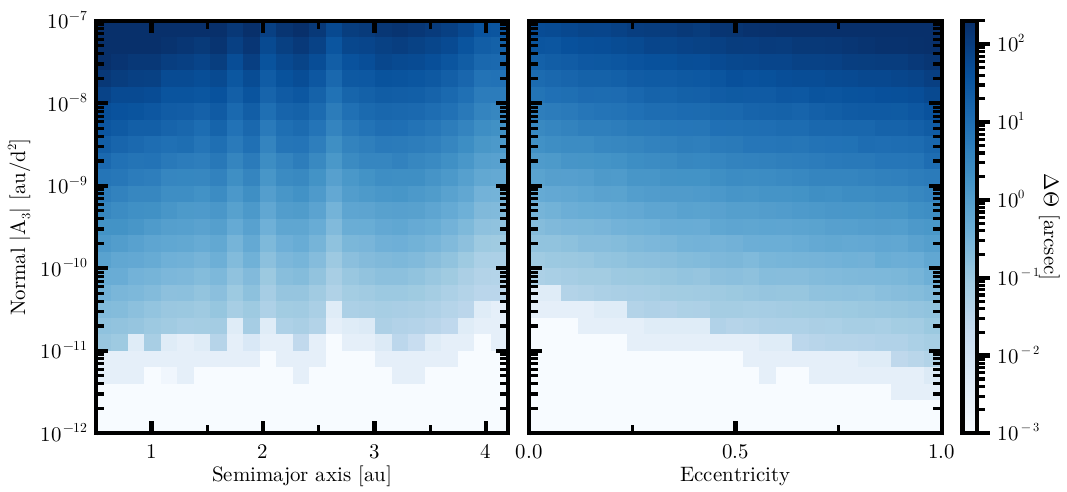}
    \vspace{-10pt}
    \caption{Similar to Fig. \ref{fig:dtheta_population}, but with out-of-plane $A_3$ acceleration instead of transverse A$_2$ acceleration. }
    \label{fig:dtheta_A3}
\end{figure*}

\begin{figure*}[!htb]
    \centering
    \includegraphics{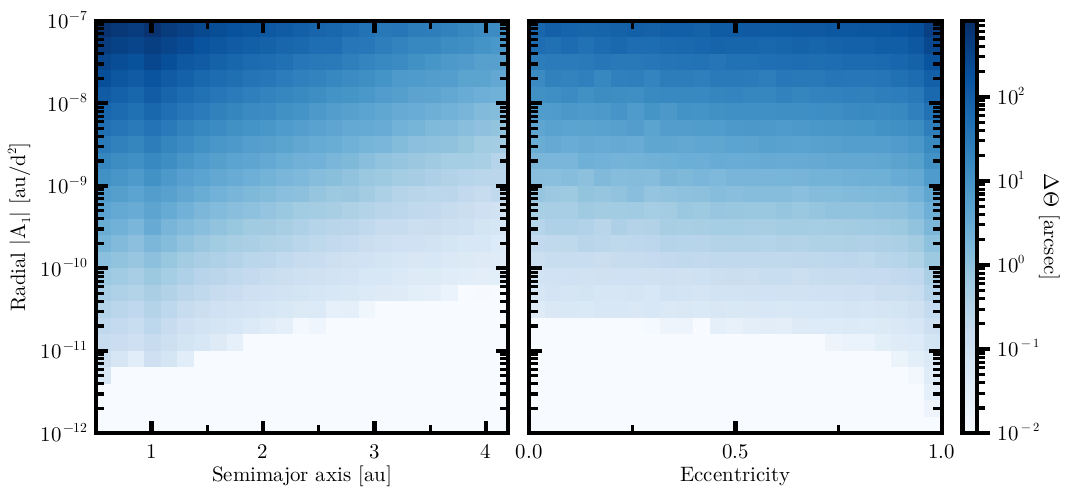}
    \vspace{-10pt}
    \caption{Similar to Fig. \ref{fig:dtheta_population}, but with radial A$_1$ acceleration and an apparition difference of 1 yr. The contour is at \qty{e3}{arcsec}.}
    \label{fig:dtheta_A1_1year}
\end{figure*}

\begin{figure*}[!htb]
    \centering
    \includegraphics{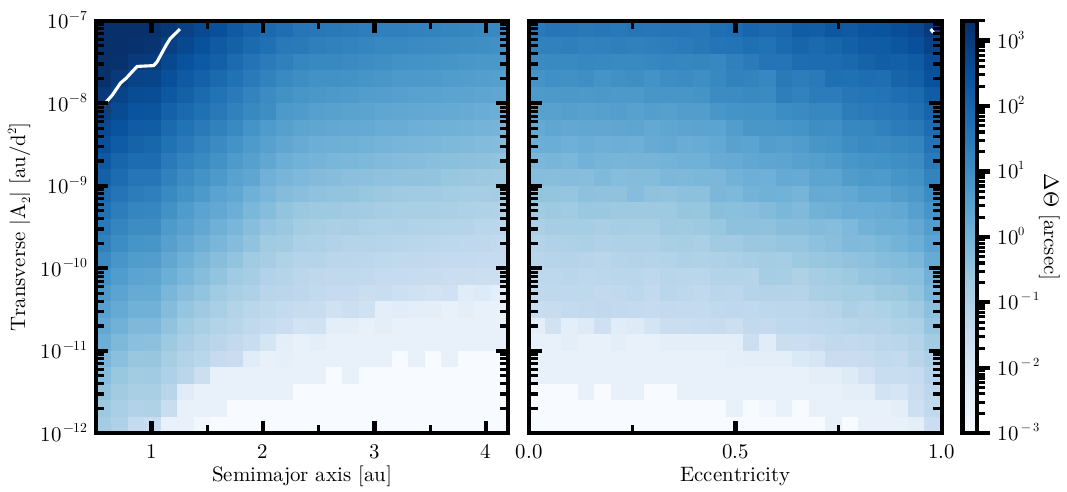}
    \vspace{-10pt}
    \caption{Similar to Fig. \ref{fig:dtheta_population}, but an apparition difference of 1 yr. The contour is at \qty{e3}{arcsec}.}
    \label{fig:dtheta_A2_1year}
\end{figure*}

\begin{figure*}[!htb]
    \centering
    \includegraphics{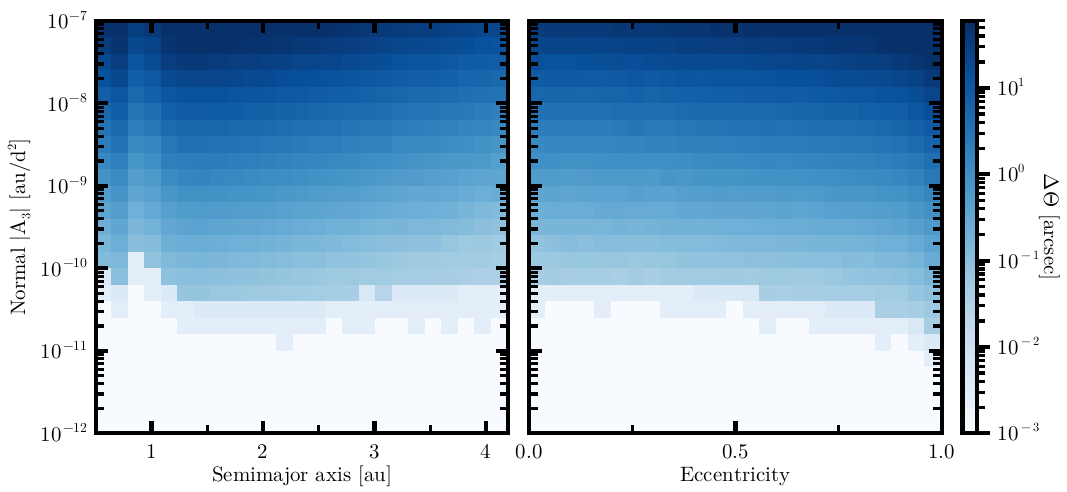}
    \vspace{-10pt}
    \caption{Similar to Fig. \ref{fig:dtheta_population}, but with out-of-plane $A_3$ acceleration and an apparition difference of \qty{1}{yr}. }
    \label{fig:dtheta_A3_1year}
\end{figure*}

\end{document}